\documentclass[11pt]{article}

\usepackage[a4paper,margin=1in]{geometry}

\usepackage{silence}
\hfuzz=\maxdimen
\vfuzz=\maxdimen
\usepackage[T1]{fontenc}
\usepackage{amsmath,amssymb,amsfonts}
\usepackage{algorithmic}
\usepackage{graphicx}
\usepackage{textcomp}
\usepackage{bm}
\usepackage{hyperref}
\usepackage[capitalize]{cleveref}
\usepackage{contour}
\usepackage{subcaption}

\usepackage[maxbibnames=2,isbn=false,url=false,doi=false,eprint=false,giveninits=true,sorting=none,style=numeric-comp]{biblatex}
\usepackage{tikz}
\usetikzlibrary{3d,decorations.pathmorphing,shapes.symbols,shapes.arrows,through}
\tikzset{positron/.style={
			->,
			blue,
			decorate,
			decoration={snake,amplitude=.75,segment length=2mm}
		},
	axis/.style={
			->,
			black!50
		}}

\usetikzlibrary{calc}
\usepackage{acro}
\acsetup{single}
\DeclareAcronym{2D}{short=2D,long=two-dimensional,first-style=short}
\DeclareAcronym{3D}{short=3D,long=three-dimensional,first-style=short}
\DeclareAcronym{CRT}{short=CRT,long=coincidence resolving time}
\DeclareAcronym{FWHM}{short=FWHM,long=full width at half maximum}
\DeclareAcronym{LFOV}{short=LFOV,long=large field-of-view}
\DeclareAcronym{MAP}{short=MAP,long=maximum a posteriori}
\DeclareAcronym{MLEM}{short=MLEM,long=Maximum-Likelihood Expectation-Maximization}
\DeclareAcronym{NM}{short=NM,long=Nelder-Mead}
\DeclareAcronym{PET}{short=PET,long=positron emission tomography}
\DeclareAcronym{Ps}{short=Ps,long=positronium}
\DeclareAcronym{o-Ps}{short=o-Ps,long=ortho-positronium}
\DeclareAcronym{p-Ps}{short=p-Ps,long=para-positronium}
\DeclareAcronym{PDF}{short=PDF,long=probability density function}
\DeclareAcronym{PLI}{short=PLI,long=positronium lifetime imaging}
\DeclareAcronym{QED}{short=QED,long=quantum electrodynamics}
\DeclareAcronym{TOF}{short=TOF,long=time-of-flight}
\DeclareAcronym{TRIO}{short=TRIO,long=Three-Photon Bayesian Imaging of Ortho-Positronium}

\newcommand{\transpose}{\ensuremath{^\intercal}}

\newcommand{\refcite}[1]{Ref.~\cite{#1}}

\newcommand{\TimeReco}{TReco}
\newcommand{\EnergyReco}{EReco}

\newcounter{csAffilNote}

\begin{document}
\title{Three-Photon Bayesian Imaging of Ortho-Positronium}
\author{L. Raczyński%
	\thanks{L. Raczyński, M. Bała, K. Klimaszewski, M. Obara, R. Y. Shopa are with the National Centre for Nuclear Research, Department of
		Complex Systems, 05-400 Otwock, Poland}%
	\setcounter{csAffilNote}{\value{footnote}}%
	, W. Krzemień\thanks{W. Krzemień is with the National Centre for Nuclear Research, Department of
		High Energy Physics, 05-400 Otwock, Poland, (e-mail:wojciech.krzemien@ncbj.gov.pl)}%
	, A. Coussat\thanks{A. Coussat is with INSA-Lyon, Université Claude Bernard Lyon 1, CNRS, Inserm, CREATIS UMR 5220, U1294, F-69373, Lyon, France}%
	, M. Bała\footnotemark[\thecsAffilNote]%
	, B.C. Hiesmayr\thanks{B. C. Hiesmayr is with the IT:U Interdisciplinary Transformation University, Freistädter Strasse 400, 4040 Linz, and University of Vienna, Faculty of Physics, Währingerstrasse 17, 1090 Vienna, Austria.}%
	\\ K. Klimaszewski\footnotemark[\thecsAffilNote]%
	, M. Obara\footnotemark[\thecsAffilNote]%
	, R. Y. Shopa\footnotemark[\thecsAffilNote]%
}
\date{}

\maketitle

\begin{abstract}
	\Ac{PET} provides functional images relying on two-photon coincidences from positron-electron annihilation. In human tissue, about 40\% of annihilations are preceded by \ac{Ps} formation, of which \ac{o-Ps} component partially decays into three photons, with the remainder annihilating via pick-off or spin-exchange into two photons. This three-photon channel carries additional information about the surrounding micro-environment, including the three-to-two-photon yield ratio as a potential diagnostic marker.
	We propose the \ac{TRIO} algorithm, a novel three-photon event-by-event image reconstruction algorithm formulated as a Bayesian maximum a posteriori inference problem. \ac{TRIO} unifies time-based trilateration, energy-based reconstruction and, for the first time, a physics-informed prior derived from the  \acl{QED} description of \ac{Ps} decay within a single probabilistic framework.
	In contrast to positronium lifetime imaging, which requires a prompt photon and is therefore restricted to specific radionuclides, \ac{TRIO} relies solely on the three photons and is fully compatible with standard radionuclides such as $^{18}$F.
	Monte Carlo simulation modelled after the Siemens Biograph Vision Quadra scanner demonstrates a mean position error of 1.65~cm, improving by a factor of about 2.8 over the time-based trilateration (4.59~cm) and by a factor of about 5.3 over energy-based reconstruction alone (8.7~cm).
	More importantly, the proposed Bayesian approach is compatible with existing \acl{TOF} \ac{PET} scanners that can register three-photon annihilation coincidences.

\end{abstract}


\acresetall

\section{Introduction}
\label{sec:introduction}
The standard \ac{PET} technique relies on detecting pairs of nearly collinear 511-keV photons produced by the annihilation of a positron, emitted by an administered radiotracer, with an electron in the surrounding medium.
By reconstructing the spatial distribution of the radiotracer concentration, a functional image of the patient's body is obtained.

Beyond the conventional two-photon annihilation paradigm, positron-electron annihilation carries additional information about its microscopic environment ~\cite{siegelPositronium1980,kacperskiThreegammaAnnihilationImaging2004, hourlierExperimentalUsesPositronium2024}. In tissues, approximately 40\% of annihilations are preceded by the formation of \ac{Ps}, a quasi-bound state of an electron and a positron~\cite{harpenPositroniumReviewSymmetry2003}. \Ac{Ps} exists in two spin configurations: \ac{p-Ps} with a lifetime of 124\,ps, which predominantly decays into two photons, and \ac{o-Ps}, which decays mainly into three photons and exhibits a mean lifetime in vacuum of 142\,ns~\cite{oreThreePhotonAnnihilationElectronPositron1949}.
The lifetime and decay characteristics of \ac{Ps} are significantly modified by the local chemical and structural environment, through additional reactions with surrounding particles, including pick-off annihilation~\cite{brandtPositroniumDecayMolecular1960, trungInvestigationOrthopositroniumAnnihilation2023} and spin-exchange interaction~\cite{ferrelOrthoParapositronium1958,zgardzinskaOrthoPara2015}. These effects make \ac{Ps}-related observables a potential source of diagnostic information beyond the conventional radiotracer spatial distribution.
This has recently motivated the extension of \ac{PET} imaging toward \ac{Ps}-based techniques~\cite{hourlierExperimentalUsesPositronium2024}.

In particular, \ac{PLI}, inspired by positron annihilation lifetime spectroscopy, has been proposed~\cite{moskalFeasibilityStudyPositronium2019b} and the first in vivo image has been delivered~\cite{moskalPositroniumImageHuman2024a}.
First results using available \ac{PET} scanners have also been reported
~\cite{huangFastHighresolutionLifetime2025,mercolliFirstPositroniumLifetime2025,mercolliPhantomImagingDemonstration2025,takyuPositroniumLifetimeMeasurement2024,huangHighResolutionPositroniumLifetime2025}.
Several algorithms dedicated to \ac{PLI} reconstruction have been proposed in recent years~\cite{qiPositroniumLifetimeImage2022,shopaPositronium2023,huangSPLITStatisticalPositronium2024, huangHighresolutionPositroniumLifetime2024, huangFastHighresolutionLifetime2025,huangStatisticalReconstruction2025}.
Despite these advances, \ac{PLI} methods share a common constraint: the prompt photon requirement restricts the technique to specific radionuclides, sometimes with limited radiopharmaceutical availability. 
Those radionuclides pose major challenges, related to positron range, sometimes low prompt branching ratio, or additional prompt-induced random and scatter components ~\cite{mercolliFirstPositroniumLifetime2025}. This motivates exploring whether the rich information content of the three-photon \ac{o-Ps} decay channel can be exploited independently, using only the annihilation photons themselves.

Unlike lifetime-based approaches, imaging the three-photon decay of \ac{o-Ps} does not require a prompt photon and is therefore compatible with standard radionuclides such as ${}^{18}\text{F}$.  Moreover, each triple coincidence encodes sufficient geometric and kinematic information to localize the annihilation vertex on an event-by-event basis.
In particular, the three-to-two photon yield ratios have been proposed as potential markers in medical imaging~\cite{kacperskiThreegammaAnnihilationImaging2004, jasinskaNewPETDiagnostic2017}, including the characterization of tissue microstructure, oxygenation~\cite{stepanovInteractionPositroniumDissolved2020}, as well as in nonmedical applications, such as characterization of porous media, and gas diffusion in materials~\cite{pevovarRatioPositronAnnihilation2007, kauppilaInvestigationsPositroniumFormation2004}.

The main challenge in three-photon imaging is the relatively low statistics of \ac{o-Ps} three-photon events compared with two-photon decays~\cite{harpenPositroniumReviewSymmetry2003, kacperskiPerformanceThreephotonPET2005, moskalFeasibilityStudyPositronium2019b}. This problem is partly mitigated by the introduction of the next-generation high-sensitivity \ac{LFOV} \ac{PET} scanners~\cite{spencerPerformanceEvaluationUEXPLORER2021a,badawiFirstHumanImaging2019,vansluisExtendingClinicalCapabilities2022, mehadjiLongAxialField2025,karakatsanisLongAxialFieldofView2026,cookImpactLongAxial2025}.
At the same time, achieving accurate localization of the three-photon annihilation vertex under realistic \acf{TOF} and energy resolution remains an unresolved problem.

Early attempts to exploit \ac{o-Ps} three-photon imaging focused on energy-based reconstruction techniques, which were derived from energy-momentum conservation~\cite{kacperskiThreegammaAnnihilationImaging2004,kacperskiPerformanceThreephotonPET2005,abuelhiaThreephotonAnnihilationPET2007}. Preliminary simulation studies and proof-of-principle experimental measurements of point sources using HP-Ge and NaI(Tl) detectors were performed.
Recently, a similar reconstruction method was applied to obtain the first image of the three-to-two decay ratio for point sources using high-resolution GAGG scintillation crystal detectors and 18F-FDG as the positron source~\cite{fujimotoAdvancingPETDirect2025}. A spatial resolution of approximately 1.1\,cm is achieved without tomographic reconstruction or \ac{TOF} information.
While conceptually appealing, these approaches require energy resolutions on the order of a few percent, which remains beyond the capabilities of current clinical PET systems.

An alternative strategy employs a GPS-like trilateration algorithm based solely on photon detection times and interaction positions~\cite{gajosTrilaterationbasedReconstructionOrthopositronium2017,moskalFeasibilityStudyPositronium2019b}. Although this method enabled the first experimental three-photon image of a three-dimensional phantom, the spatial resolution is limited to approximately 8 cm~\cite{moskalTestingCPTSymmetry2021b}.

All aforementioned methods implicitly assume isotropic photon emission, neglecting the correlations imposed by the \ac{QED} description of \ac{o-Ps} decay~\cite{beresteckijQuantumElectrodynamics2008} (see also ~\cite{raczynskiVertex2026}). Therefore, a method that combines timing, energy, and fundamental physics constraints within a unified reconstruction framework is required to fully exploit three-photon annihilation for practical imaging.

In this work, we propose the \ac{TRIO} technique, a novel three-photon event-by-event image reconstruction framework formulated as a Bayesian \ac{MAP} inference problem. TRIO unifies all available information within a single probabilistic model by incorporating both time-based trilateration and energy-based reconstruction, including, for the first time, constraints from physics modelling.
This Bayesian approach enables a substantial improvement in spatial resolution compared to state-of-the-art timing-only or energy-only methods, while relaxing the stringent energy resolution requirements of earlier approaches.
More importantly, the proposed algorithm is compatible with existing \ac{TOF}-\ac{PET} detector scanners, provided that the scanners can detect photons at lower energies (below 511\,keV) and that they are configured to register three-photon coincidences.

This works focuses on evaluation of the \ac{TRIO} algorithm itself and voluntarily does not attempt at assessing its clinical applicability. More specifically, many effects of \ac{PET} physics are purposefully ignored, including detector scattering, phantom scattering, random coincidences or attenuation, in order to focus on raw performance of the \ac{TRIO} algorithms on idealized data. As a consequence, the emphasis was put on the impacts of time and energy resolutions of the scanner on the resulting three-gamma \ac{Ps} image, and assessment of the impact of other type of physical effects are left for future work.

The remainder of the article is organized as follows:
\cref{sec:ortho-reco} presents the TRIO algorithm and reviews existing methods.
\Cref{sec:MC} describes the Monte Carlo simulation model and evaluation metrics.
\Cref{sec:results} presents the results, followed by a discussion in \cref{sec:discussion}. Conclusions are given in \cref{sec:conclusion}.

\section{\Acl*{o-Ps} position reconstruction}
\label{sec:ortho-reco}

\begin{figure}
	\centering
	\begin{subfigure}[t]{.49\textwidth}
		\centering
		\begin{tikzpicture}
			\def\PETLength{5}
			\def\PETRadius{3}
			\def\PETAngleStep{12}
			\foreach \a in {0,\PETAngleStep,...,360}
				{
					\draw[gray!50] (xyz cylindrical cs:radius=\PETRadius,angle=\a,z=-\PETLength/2)
					-- (xyz cylindrical cs:radius=\PETRadius,angle=\a+\PETAngleStep,z=-\PETLength/2)
					-- ++(0,0,\PETLength)
					-- (xyz cylindrical cs:radius=\PETRadius,angle=\a,z=\PETLength/2) ;
				}

			\draw[axis] (0,0,0) -- (2.4,0,0) node [right] {$x'$} ;
			\draw[axis] (0,0,0) -- (0,2.4,0) node [above] {$y'$} ;
			\draw[axis] (0,0,0) -- (0,0,3.2) node [below left] {$z'$} ;

			\coordinate (A) at (.5,.5,2.5) ;
			\coordinate (H1) at (2.9992866403939344,-0.06541902440777818,2.907689130575759) ;
			\coordinate (H2) at (-2.969993249340852,-0.4232494522970675,0.8049153025040762) ;
			\coordinate (H3) at (-0.8868778727010976,2.865911310371132,3.630103246796525) ;
			\draw[positron] (A)	-- (H1) node [above] {\contour{white}{$\gamma_1$}} node [below] {\contour{white}{$(\mathbf{x}'_1,E_1,t_1)$}} ;
			\draw[positron] (A) -- (H2) node [above,left] {\contour{white}{$\gamma_2$}} node [below] {\contour{white}{$(\mathbf{x}'_2,E_2,t_2)$}} ;
			\draw[positron] (A) -- (H3) node [above] {\contour{white}{$\gamma_3$}} node [below] {\contour{white}{$(\mathbf{x}'_3,E_3,t_3)$}} ;
			\draw[black, thick] (H1) -- (H2) -- (H3) -- cycle ;

			\node[draw=none,starburst,fill=red,starburst point height=1,inner sep=1pt,label={[red]above:\contour{white}{\acs*{o-Ps} decay}},label={[red]below:\contour{white}{$(\mathbf{x}',E,T)$}}] at (A) {} ;
		\end{tikzpicture}
		\caption{}
		\label{subfig:oPsAnnihilation}
	\end{subfigure}
	\hfill
	\\
	\begin{subfigure}[t]{.49\textwidth}
		\centering
		\begin{tikzpicture}
			\draw[axis] (-3.5,0) -- (3.5,0) node [right] {$x$} ;
			\draw[axis] (0,-1.5) -- (0,3.5) node [above] {$y$} ;

			\def\clipSize{3.8}
			\def\clipSizeYmin{-1.5}
			\clip (-\clipSize,\clipSizeYmin) rectangle (\clipSize,\clipSize) ;

			\coordinate (A) at (0.5,0.5) ;
			\coordinate (H1) at (3.0,0.0) ;
			\coordinate (H2) at (-2.9344441518506557,-0.6237287228190531) ;
			\coordinate (H3) at (-0.31359707966583783,2.983564457427568) ;
			\draw[positron] (A)	-- (H1) node[above,midway] {$r_1$}   node [below] {\contour{white}{$(\mathbf{x}_1,E_1,t_1)$}} ;
			\node [draw,dashed,gray] at (H1) [circle through={(A)}] {};
			\draw[positron] (A) -- (H2)
			node[above,midway] {$r_2$}
			node [below] {\contour{white}{$(\mathbf{x}_2,E_2,t_2)$}} ;
			\node [draw,dashed,gray] at (H2) [circle through={(A)}] {};
			\draw[positron] (A) -- (H3)
			node[right,midway] {$r_3$}
			node [above] {\contour{white}{$(\mathbf{x}_3,E_3,t_3)$}} ;
			\node [draw,dashed,gray] at (H3) [circle through={(A)}] {};


			\draw[black, thick] (H1)  -- (H2) node[below,midway] {$d_{12}$};
			\draw[black, thick] (H2) --  (H3) node[left,midway] {$d_{23}$};
			\draw[black, thick] (H1) --  (H3) node[right,midway] {$d_{13}$};
			\node[draw=none,starburst,fill=red,starburst point height=1,inner sep=1pt] at (A) {}; 

			\def\angleRadius{0.75}
			\draw[blue] let \p1=(A), \p2=(H1), \p3=(H2),
			\n1={atan2(\y2-\y1,\x2-\x1)}, \n2={atan2(\y3-\y1,\x3-\x1)} in
			($(A)+(\n1:\angleRadius)$) arc (\n1:\n2:\angleRadius);
			\draw[blue] let \p1=(A), \p2=(H1), \p3=(H2),
			\n1={atan2(\y2-\y1,\x2-\x1)}, \n2={atan2(\y3-\y1,\x3-\x1)} in
			($(A)+(0.5*\n1+0.5*\n2:\angleRadius*0.5)$) node[anchor=center]{$\alpha_{12}$};

			\draw[blue] let \p1=(A), \p2=(H2), \p3=(H3),
			\n1={atan2(\y2-\y1,\x2-\x1)}, \n2={atan2(\y3-\y1,\x3-\x1)} in
			($(A)+(\n1:\angleRadius)$) arc (\n1:\n2:\angleRadius);
			\draw[blue] let \p1=(A), \p2=(H2), \p3=(H3),
			\n1={atan2(\y2-\y1,\x2-\x1)}, \n2={atan2(\y3-\y1,\x3-\x1)} in
			($(A)+(0.5*\n1+0.5*\n2-180:\angleRadius*0.65)$) node[anchor=center]{$\alpha_{23}$};

			\draw[blue] let \p1=(A), \p2=(H3), \p3=(H1),
			\n1={atan2(\y2-\y1,\x2-\x1)}, \n2={atan2(\y3-\y1,\x3-\x1)} in
			($(A)+(\n1:\angleRadius)$) arc (\n1:\n2-360:\angleRadius);
			\draw[blue] let \p1=(A), \p2=(H1), \p3=(H3),
			\n1={atan2(\y2-\y1,\x2-\x1)}, \n2={atan2(\y3-\y1,\x3-\x1)} in
			($(A)+(0.5*\n1+0.5*\n2:\angleRadius*0.5)$) node[anchor=center]{$\alpha_{13}$};

		\end{tikzpicture}
		\caption{}
		\label{fig:oPs2D}
	\end{subfigure}
	\caption{Example of a three-photon
		\acs*{o-Ps} annihilation. (a) Schematic 3D representation in the Quadra  Siemens scanner (b) Same annihilation transformed to a \acs*{2D} plane, possible due to the co-planarity of the four points.}
	\label{fig:oPs3D2D}
\end{figure}
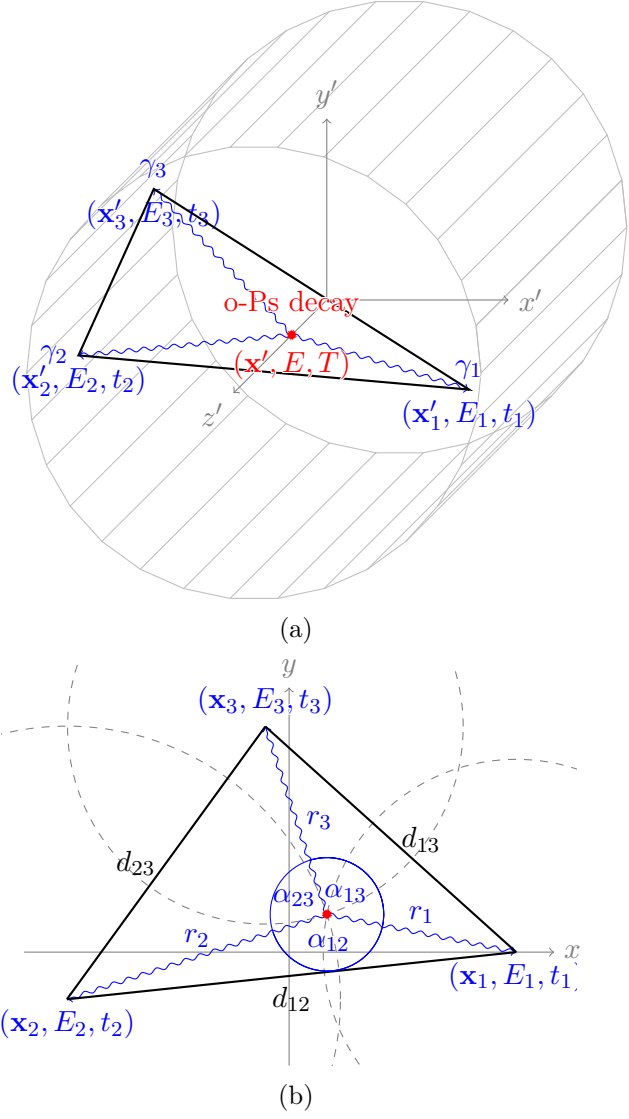

In this section, we introduce a Bayesian-based \ac{o-Ps} event position reconstruction method.
Each \ac{o-Ps} event is represented by three hit detection positions, times and energies.
The conservation of momentum implies that the hit detection positions and the annihilation position all lie on the same plane (see e.g.~\cite{raczynskiVertex2026}), therefore, the original \ac{3D} measurement space $(x',y',z')$ may be transformed to search for the solution in the \ac{2D} decay plane ($x,y$).
The transformed coordinates of the interaction positions will be denoted with $\mathbf{x}_i \triangleq ({x_i}, {y_i})\transpose$, where $i =1, 2, 3$. A summary of the notation used in the following can be found in \cref{fig:oPs3D2D}.

The \ac{o-Ps} decay position may be evaluated using a time-based trilateration model or an energy-based reconstruction.

In the time-based trilateration model, the decay position is  calculated using the hit detection positions $\mathbf{x}_i$ and times $t_i$ (for $i =1, 2, 3$) as an intersection of three circles with centers at $\mathbf{x}_i$ and radii equal to the product of the photon's time of flight and the speed of light (see \cref{fig:oPs2D}): 
\begin{equation}
	\label{eq:gajosTrilateration}
	\begin{cases}
		(x_t-x_1)^2+(y_t-y_1)^2=c^2(t_1-T)^2 \\
		(x_t-x_2)^2+(y_t-y_2)^2=c^2(t_2-T)^2 \\
		(x_t-x_3)^2+(y_t-y_3)^2=c^2(t_3-T)^2
	\end{cases}.
\end{equation}
The solution of the above system of equations yields the \ac{o-Ps} decay position ($\mathbf{x_t} \triangleq (x_t, y_t)\transpose$) and time ($T$).
At maximum, two intersections can exist, but only one with $T \leq \min(t_1, t_2, t_3)$ is physical, as the decay time precedes all the registration times. 
The position $\mathbf{x_t}$ is transformed back into the \ac{3D} space, yielding the reconstructed vertex. Details of this algorithm can be found in \refcite{gajosTrilaterationbasedReconstructionOrthopositronium2017}.
Since the analytical solution of the trilateration problem exists, the reconstructed position $\mathbf{x_t}$ may be written as a function:
\begin{equation}
	\mathbf{x_t} = g_t \left(\mathbf{x}_1, \mathbf{x}_2,  \mathbf{x}_3, \mathbf{t} \right)
	\label{Eq_gA}
\end{equation}
where $\mathbf{t} = (t_1, t_2, t_3)\transpose$ is the vector that holds the interaction times.

In the energy-based approach, the decay position is derived using the hit detection positions ($x_i, y_i$) and energies $E_i,$   where $i =1, 2, 3$.
The algorithm proposed by Kacperski et al.~\cite{kacperskiThreegammaAnnihilationImaging2004}
exploits the momentum conservation law to define that
\begin{equation}
	\label{eq:kacperskiMomentum}
	\begin{cases}
		\frac{E_1}{c} \frac{x_E - x_1}{\lVert \mathbf{x_E} - \mathbf{x}_1 \rVert}
		+ \frac{E_2}{c} \frac{x_E - x_2}{\lVert \mathbf{x_E} - \mathbf{x}_2 \rVert}
		+ \frac{E_3}{c} \frac{x_E - x_3}{\lVert \mathbf{x_E} - \mathbf{x}_3 \rVert}
		= 0 \\
		\frac{E_1}{c} \frac{y_E - y_1}{\lVert \mathbf{x_E} - \mathbf{x}_1 \rVert}
		+ \frac{E_2}{c} \frac{y_E - y_2}{\lVert \mathbf{x_E} - \mathbf{x}_2 \rVert}
		+ \frac{E_3}{c} \frac{y_E - y_3}{\lVert \mathbf{x_E} - \mathbf{x}_3 \rVert}
		= 0
	\end{cases}
\end{equation}
where $\mathbf{x_E} \triangleq (x_E, y_E)\transpose$.
The non-linear system of equations in \cref{eq:kacperskiMomentum} can be solved
iteratively or analytically under the constraints from the energy conservation law:
\begin{equation}
	E_1 + E_2 + E_3 = 2 m_e c^2 \approx 1022\;\mathrm{keV}.
\end{equation}
The authors of the original work do not describe how the system of equations was solved. The detailed derivation of the exact solution for the energy-based model can be found in ~\cite{raczynskiVertex2026}.
Therefore, the reconstructed position can be given as a function:
\begin{equation}
	\mathbf{x_E} = g_E \left(\mathbf{x}_1, \mathbf{x}_2,  \mathbf{x}_3, \mathbf{E} \right),
	\label{Eq_gB}
\end{equation}
where $\mathbf{E} = (E_1, E_2, E_3)\transpose$ is the vector that holds the energies of the annihilation photons.

It is important to emphasize that, given an ideal detection system, the calculated values $\mathbf{x_t}$ and $\mathbf{x_E}$, in \cref{Eq_gA,Eq_gB}, respectively, would be identical and would both correspond to the exact position of the \ac{o-Ps} decay.
However, in practice, the measured position, time and energy information is known to a certain precision, and instead of vectors $\mathbf{x}_1, \mathbf{x}_2,  \mathbf{x}_3, \mathbf{t}, \mathbf{E},$ corresponding random variables $\mathbf{\hat{x}}_1, \mathbf{\hat{x}}_2,  \mathbf{\hat{x}}_3, \mathbf{\hat{t}}, \mathbf{\hat{E}}$ have to be considered.
Consequently, the calculations provided by \cref{Eq_gA,Eq_gB} yield estimators $\mathbf{\hat{x}_t}$ and $\mathbf{\hat{x}_E}$, respectively.

In the proposed approach, the random variables $\mathbf{\hat{x}_t}$ and $\mathbf{\hat{x}_E}$  are combined using the Bayesian framework.
The unknown decay position is estimated by maximizing the posterior \ac{PDF} of the decay position $\mathbf{x}$ conditional on both measurements
$\mathbf{\hat{x}_t}$ and $\mathbf{\hat{x}_E}$, namely $P(\mathbf{x} | \mathbf{x_t} \cap \mathbf{x_E}).$
Using the Bayesian approach, the conditional \ac{PDF} on a common distribution may be calculated as:
\begin{equation}
	P(\mathbf{x} | \mathbf{x_t} \cap \mathbf{x_E}) =  \frac{P(\mathbf{x_t} \cap \mathbf{x_E} | \mathbf{x})\,P(\mathbf{x})}{P(\mathbf{x_t} \cap \mathbf{x_E})}, \label{Eq_P_x_xt_xE_1}
\end{equation}
where $P(\mathbf{x})$ is the prior distribution of decay position $\mathbf{x}$ and
the probability $P(\mathbf{x_t} \cap \mathbf{x_E})$ is independent of $\mathbf{x}$, and  serves as a normalization constant.
Assuming conditional independence of random variables $\mathbf{\hat{x}_t}$ and $\mathbf{\hat{x}_E}$ on decay position $\mathbf{x}$,
the likelihood of common reconstruction $P(\mathbf{x_t} \cap \mathbf{x_E} | \mathbf{x})$ is given as:
\begin{equation}
	P(\mathbf{x_t} \cap \mathbf{x_E} | \mathbf{x}) = P(\mathbf{x_t}|\mathbf{x}) \, P(\mathbf{x_E}|\mathbf{x}),
\end{equation}
and is a product of the likelihoods $P(\mathbf{x_t} | \mathbf{x})$ and $P(\mathbf{x_E} | \mathbf{x})$ of reconstructions $\mathbf{\hat{x}_t}$ and $\mathbf{\hat{x}_E},$ respectively.
Therefore, the conditional probability in \cref{Eq_P_x_xt_xE_1} reads:
\begin{equation}
	P(\mathbf{x} | \mathbf{x_t} \cap \mathbf{x_E}) \propto \,P(\mathbf{x_t} | \mathbf{x}) \, P(\mathbf{x_E} | \mathbf{x})\, P(\mathbf{x})
\end{equation}
and finally, the decay position evaluated based on the Bayesian approach, denoted hereafter with $\mathbf{\hat{x}_B}$, is:
\begin{equation}
	\mathbf{\hat{x}_B} = \arg \max \left\{  P(\mathbf{x_t} | \mathbf{x}) \, P(\mathbf{x_E} | \mathbf{x})\, P(\mathbf{x}) \right\}.		\label{Eq_Bayes_xAxB}
\end{equation}

We wish to comment on the
assumption that the time and energy reconstructions are conditionally independent.
The uncertainties related to the registered times ($\mathbf{\hat{t}}$) and energies ($\mathbf{\hat{E}}$) are independent.
These random variables are used to calculate the random positions $\mathbf{\hat{x}_t}$ and $\mathbf{\hat{x}_E}$ based on \cref{Eq_gA,Eq_gB}, respectively. When the  interaction position measurement $\mathbf{x}_1, \mathbf{x}_2,  \mathbf{x}_3$ errors are small with respect to the time (energy) uncertainty, the
probability distribution related
to the reconstruction  $\mathbf{\hat{x}_t}$ ($\mathbf{\hat{x}_E}$) may be considered exclusively in time (energy) space, and  $\mathbf{\hat{x}_t}$ and $\mathbf{\hat{x}_E}$ are conditionally independent. In general, the higher the position measurement uncertainty, the weaker the assumption of the conditional independence of  $\mathbf{\hat{x}_t}$ and $\mathbf{\hat{x}_E}$.

Note that in the case of  separate reconstruction of time-based and energy-based \ac{o-Ps} decay position, the estimators $\mathbf{\hat{x}_t}$ and $\mathbf{\hat{x}_E}$
calculated according to \cref{Eq_gA,Eq_gB}, respectively, are maximum likelihood estimators, that is:
\begin{align}
	\mathbf{\hat{x}_t} & =g_t \left(\mathbf{x}_1, \mathbf{x}_2,  \mathbf{x}_3, \mathbf{\hat{t}} \right) = \arg \max P(\mathbf{x_t} | \mathbf{x})  \label{Eq_Bayes_xA}   \\
	\mathbf{\hat{x}_E} & = g_E \left(\mathbf{x}_1, \mathbf{x}_2,  \mathbf{x}_3,  \mathbf{\hat{E}} \right) = \arg \max P(\mathbf{x_E} | \mathbf{x}). \label{Eq_Bayes_xB}
\end{align}
In contrast to the proposed Bayesian approach in \cref{Eq_Bayes_xAxB}, the evaluations based on functions $g_t$ and $g_E$  do not additionally take into account
a physics-driven prior \ac{PDF} $P(\mathbf{x})$  derived from the QED-compliant model of the decay of three-photon \ac{o-Ps}.
The algorithms performing the independent time- and energy-based calculations according to \cref{Eq_Bayes_xA,Eq_Bayes_xB},
respectively,
were implemented for the comparative study with the proposed method.

It should be stressed that in the case of the \ac{TRIO} method, the global maximum of the function in \cref{Eq_Bayes_xAxB} exists but cannot be easily found using the classical derivative-based optimization methods due to its non-trivial form.
For this reason, a derivative-free \ac{NM} method~\cite{nelderSimplexMethodFunction1965} is applied.
The idea behind the NM algorithm is to move toward the optimum by performing basic operations: reflection, expansion, contraction, and shrinking. 
It is demonstrated that the NM method converges well for solving small-dimensional problems~\cite{lagariasConvergencePropertiesNelderMead1998,hanEffectDimensionalityNelder2006}, and was successfully applied in recent image reconstruction studies~\cite{shopaOptimisation2021,shopaPositronium2023}.

To apply the NM method for finding the maximum of the function in \cref{Eq_Bayes_xAxB}, the \ac{PDF}s $P(\mathbf{x_t} | \mathbf{x}), P(\mathbf{x_E} | \mathbf{x})$ and $P(\mathbf{x})$ must be determined.
Those three probability components need to be evaluated at each point  ($x,y$) in the decay plane indicated by the NM algorithm during the iterative search.
For this purpose, in the following, the derivation of three probability distributions in \cref{Eq_Bayes_xAxB} are presented.

\subsection{Derivation of probability distribution \texorpdfstring{$P(\mathbf{x_t} | \mathbf{x})$}{P(xt|x)}} \label{sec:P_xT}

For each point $\mathbf{x_t}$ in decay plane, the probability $P(\mathbf{x_t} | \mathbf{x})$ may be evaluated using the measured interaction positions $\mathbf{\hat{x}}_i$ and times  $\hat{t}_i$ for  $i = 1, 2, 3$ (represented as circles of radius $c(t_i-T)$ in \cref{fig:oPs2D}).
The idea behind estimation of the \ac{PDF} $P(\mathbf{x_t} | \mathbf{x})$ is that
a hypothetical registration time $\mathbf{\tilde{t}} = (\tilde{t}_1, \tilde{t}_2, \tilde{t}_3)\transpose$ related to the decay point  $\mathbf{x_t}$ by the function
\begin{equation}
	\mathbf{\tilde{t}} = h_t \left(\mathbf{x}_1, \mathbf{x}_2,  \mathbf{x}_3, \mathbf{x_t} \right),
	\label{Eq_h_T}
\end{equation}
may be  evaluated and compared to the measured times $\mathbf{\hat{t}}= (\hat{t}_1, \hat{t}_2, \hat{t}_3)\transpose$.
We assume that the times $\hat{t}_i$ are normally distributed with mean $t_i$  and standard deviation $\sigma_t,$ independent of the value $\hat{t}_i$,
and that the uncertainties associated with interaction position measurements are negligible, that is: $\mathbf{\hat{x}}_i = \mathbf{x}_i$ for  $i = 1, 2, 3.$
Therefore, the probability distribution related to the measurement space may be considered exclusively in time space, and
the value of probability $P(\mathbf{x_t} | \mathbf{x})$ is  given as:
\begin{equation}
	P(\mathbf{x_t} | \mathbf{x}) = \frac{1}{\sqrt{(2\pi)^3 |\mathbf{\Sigma}_t|}} e^{ -\frac{1}{2}\left(\mathbf{\hat{t}} - \mathbf{\tilde{t}}(\mathbf{x_t})\right)\transpose \mathbf{\Sigma}_t^{-1}\left(\mathbf{\hat{t}} - \mathbf{\tilde{t}}(\mathbf{x_t})\right)},
	\label{Eq_P_xt_x_NORMAL}
\end{equation}
where $\mathbf{\Sigma}_t =\sigma_t^2 \mathbf{I}$ is covariance matrix.

The function $h_t$ defined in \cref{Eq_h_T}, translating the information from a given  point $\mathbf{x_t}$ in decay plane to a hypothetical registration time $\mathbf{\tilde{t}}$ can be calculated as:
\begin{equation}
	\tilde{t}_i = {T} + \Delta t_i = {T} + \frac{r_i}{c}, \label{Eq_tA_tilde}
\end{equation}
where ${T}$ is unknown decay time corresponding to selected  $\mathbf{x_t}$ and $r_i$ are the distances between $\mathbf{x_t}$ and three interaction positions:
\begin{equation}
	r_i = \lVert\mathbf{x}_i - \mathbf{x_t}\rVert, \label{Eq_ri}
\end{equation}
for $i = 1, 2, 3.$
The value of decay time  may be found as the maximum log-likelihood estimate, that is, after maximizing a logarithm of the probability in \cref{Eq_P_xt_x_NORMAL}, i.e.,
\begin{equation}
	\tilde{T} = \arg \max \left\{ -\frac{1}{2}\left(\mathbf{\hat{t}} - \mathbf{\tilde{t}}({T})\right)\transpose \mathbf{\Sigma}_t^{-1}\left(\mathbf{\hat{t}} - \mathbf{\tilde{t}}({T})\right) + \nu \right\}
	\label{Eq_log_likelihood}
\end{equation}
where $\mathbf{\tilde{t}}({T})$ is defined by \cref{Eq_tA_tilde} and the last term $\nu = -\frac{1}{2}\ln\left((2\pi)^3 |\mathbf{\Sigma}_t|\right)$
is independent of $T.$
The log-likelihood in \cref{Eq_log_likelihood} is a quadratic function of $T$, and the closed-form solution of decay time is given by:
\begin{equation*}
	\tilde{T} = \frac{1}{3} \left(\sum_{i=1}^3 \hat{t}_i -  \frac{r_i}{c} \right)
\end{equation*}
and is an average of the decay times given by the three interaction positions.
Hence, the hypothetical registration times in \cref{Eq_tA_tilde} related to the decay point $\mathbf{x_t}$ are given as:
\begin{equation}
	\tilde{t}_i = \frac{r_i}{c} + \frac{1}{3} \left(\sum_{j=1}^3 \hat{t}_j -  \frac{r_j}{c} \right)
	\label{Eq_hypo_time}
\end{equation}
and these values are substituted to \cref{Eq_P_xt_x_NORMAL} to evaluate the final value of the conditional probability
$P(\mathbf{x_t} | \mathbf{x}).$

\subsection{Derivation of probability distribution \texorpdfstring{$P(\mathbf{x_E} | \mathbf{x})$}{P(xE|x)}}
\label{sec:P_xE}

For each point $\mathbf{x_E}$ in decay plane, the probability $P(\mathbf{x_E} | \mathbf{x})$ may be evaluated using the measured interaction positions $\mathbf{\hat{x}}_i$ and energies  $\hat{E}_i$ for  $i = 1, 2, 3.$
The idea behind estimation of the \ac{PDF} $P(\mathbf{x_E} | \mathbf{x})$ is that
a hypothetical energy vector
$\mathbf{\tilde{E}} = (\tilde{E}_1, \tilde{E}_2, \tilde{E}_3)\transpose$ related to the decay point  $\mathbf{x_E}$ by the function
\begin{equation}
	\mathbf{\tilde{E}} = h_E \left(\mathbf{x}_1, \mathbf{x}_2,  \mathbf{x}_3, \mathbf{x_E} \right),
	\label{Eq_h_E}
\end{equation}
may be  evaluated and compared to the measured energies
$\mathbf{\hat{E}}= (\hat{E}_1, \hat{E}_2, \hat{E}_3)\transpose$.
As before, we assume that the uncertainties associated with interaction position measurements are negligible, that is: $\mathbf{\hat{x}}_i = \mathbf{x}_i.$
On the other hand, the energies $\hat{E}_i$ are assumed to be normally distributed with mean $E_i$  and standard deviation $\sigma_i:$
\begin{equation}
	\sigma_i = \eta \sqrt{\hat{E}_i},
	\label{Eq_energy_res}
\end{equation}
where $\eta$ is a scanner-specific parameter describing the energy resolution.
Therefore, the probability distribution related to the measurement space may be considered exclusively in the energy space, and the value of probability $P(\mathbf{x_E} | \mathbf{x})$ is given as:
\begin{equation}
	P(\mathbf{x_E} | \mathbf{x}) = \frac{1}{\sqrt{(2\pi)^3 |\mathbf{\Sigma}_E|}} e^{-\frac{1}{2}\left(\mathbf{\hat{E}} - \mathbf{\tilde{E}}(\mathbf{x_E})\right)\transpose \mathbf{\Sigma}_E^{-1}\left(\mathbf{\hat{E}} - \mathbf{\tilde{E}}(\mathbf{x_E})\right)},
	\label{Eq_P_x_xE_NORMAL}
\end{equation}
where the covariance matrix:
\begin{equation}
	\mathbf{\Sigma}_E = \begin{bmatrix}
		\sigma_1^2 & 0          & 0          \\
		0          & \sigma_2^2 & 0          \\
		0          & 0          & \sigma_3^2
	\end{bmatrix}.
	\label{Eq_cov_Sigma_E}
\end{equation}

The evaluation of the function $h_E$ in \cref{Eq_h_E}, translating the information from the decay plane to the energy space, requires several steps.
First, for a given  point $\mathbf{x_E}$ in decay plane, the opening angles  $\alpha_{ij}$ (see \cref{fig:oPs2D} for details) are calculated using the cosine rule:
\begin{equation}
	\label{Eq_alpha_ij_cosine_rule}
	\alpha_{ij} = \arccos \left(\frac{r_i^2 + r_j^2 - d_{ij}^2}{2r_i r_j} \right)  \quad\quad 1 \leq i < j \leq 3
\end{equation}
where the distances $r_i$ are evaluated as shown in \cref{Eq_ri} and $d_{ij} = \lVert\mathbf{x}_i - \mathbf{x}_j\rVert$.
The angles in the triangle defined by the momentum vectors  are given by~\cite{raczynskiVertex2026}:
\begin{equation}
	\theta_{ij} = \pi - \alpha_{ij}	\quad\quad 1 \leq i < j \leq 3.
\end{equation}
Finally, using the sine rule, the momentum elements may be expressed as a function of angles $\theta_{ij}.$
As the energies of the annihilation photons are proportional to their momentum vectors, a hypothetical energy related to the decay point
$\mathbf{x_E}$ are given as:
\begin{equation}
	\label{Eq_omega_i}
	\tilde{E}_i = \frac{2 m_e c^2 \sin\theta_{jk}}{\sum_{j'k'} \sin\theta_{j'k'} }, \quad k \neq j\neq i.
\end{equation}
These values are substituted into \cref{Eq_P_x_xE_NORMAL} to evaluate the conditional probability
$P(\mathbf{x_E} | \mathbf{x}).$

\subsection{Derivation of prior probability distribution \texorpdfstring{$P(\mathbf{x})$}{P(x)}}
\label{sec:P_x}

As it is shown in \cref{sec:P_xE}, for a given point $\mathbf{x}$ in decay plane,  a hypothetical energy vector $\mathbf{\tilde{E}} = (\tilde{E}_1, \tilde{E}_2, \tilde{E}_3)\transpose$ may be calculated.
The prior probability distribution $P(\mathbf{x})$ encodes the physical constraints imposed by the QED description of the \ac{o-Ps} decay, independently of the measured times and energies. Since the emission angles - and therefore the hypothetical energies are derived deterministically by the candidate position $\mathbf{x}$ alone, this defines a smooth deterministic map $\mathbf{x} \rightarrow \tilde{E}(\mathbf{x})$, which allows to transform the QED probability distribution over photon momenta to a corresponding distribution over spatial positions.

It should be noted that under the  assumption that the uncertainties associated with interaction position measurements are negligible, that is: $\mathbf{\hat{x}}_i = \mathbf{x}_i,$
the $\mathbf{\tilde{E}}$ in \cref{Eq_omega_i} corresponds to the exact value of the energies related to the decay point $\mathbf{x}.$
In previous section \cref{sec:P_xE} it is demonstrated that the vector $\mathbf{\tilde{E}}$ can be compared with the measured energies $\mathbf{\hat{E}}$ and as a result the probability
$P(\mathbf{x_E} | \mathbf{x})$ can be evaluated.
In this section, the prior probability distribution of  $\mathbf{x},$ calculated in the absence of the energy ($\mathbf{\hat{E}}$) and time ($\mathbf{\hat{t}}$) information,
is introduced by explicitly incorporating the correlations between photon energies and emission angles imposed by QED physics.

It can be derived from~\cite{oreThreePhotonAnnihilationElectronPositron1949,beresteckijQuantumElectrodynamics2008}  that not every combination of the momentum vectors is equally probable.
The \ac{PDF}  in the momentum space and equivalently in the energy space can be expressed as:
\begin{equation}
	P({\mathbf{\tilde{E}}}(\mathbf{x_E})) =  \kappa \sum_{i=1}^3\left(\frac{m_e c^2 -\tilde{E}_i}{\tilde{E}_j \tilde{E}_k}\right)^2, \quad k \neq j\neq i
	\label{Eq_Prob_E_priori}
\end{equation}
where hypothetical energy vector ($\mathbf{\tilde{E}}$) is calculated exactly the same as shown in \cref{sec:P_xE} in \cref{Eq_omega_i} and   $\kappa$ is a normalization constant. For clarity of the description the relation $\tilde{E}_i = \tilde{E}_i(\mathbf{x_E})$ in right-hand side in \cref{Eq_Prob_E_priori}, was omitted.
Finally, the value of probability $P(\mathbf{x})$ is given as:
\begin{equation}
	P(\mathbf{x}) = P({\mathbf{\tilde{E}}}(\mathbf{x_E})) .
	\label{Eq_Prob_x}
\end{equation}
It is important to note that this prior is evaluated using the geometrically predicted energies $\mathbf{\tilde{E}}$, not the measured energies $\mathbf{\hat{E}}$. It therefore represents a pure physical constraint on which decay positions are kinematically favoured, before energy measurements are taken into account. As a consequence, $P(\mathbf{x})$ and the $P(\mathbf{x_E}|\mathbf{x})$ in \cref{Eq_Bayes_xAxB} do not double-count the energy information; the former depends only on the geometry of the candidate position, while the latter compares the predicted energies against the actual measurements.

\section{Monte Carlo simulations and algorithms implementations}
\label{sec:MC}
We tested the algorithm's performance by conducting Monte Carlo simulation studies of a \acs{TOF}-\acs{PET} detector with sources emitting three photons originating from positronium decays.
The simulations are performed by the dedicated Python library
developed in the frame of the EuroHPC-PL project and further extended for the current studies. The simulation package includes the \ac{o-Ps} decay model compliant with the \ac{QED} description of the positronium decay~\cite{beresteckijQuantumElectrodynamics2008}, which provides the kinematics of the emitted photons, and the transport code calculates the intersection of the photon trajectories with the scanner material. It is assumed that the photons interact with the detector material only via the photoelectric effect.

The physics model of the package is validated by a series of tests and benchmarks, including physics-oriented tests reproducing the expected dependencies between the kinematic variables, their probability distributions, as well as the fulfilment of physics-imposed constraints such as momentum-energy conservation.

In the current study, we simulated a cylindrical PET scanner with a radius of 41~cm and an axial length of 106~cm. The system consisted of one scintillator layer made of 322 rings, each ring containing 760 detectors with a size of $3.2\times 3.2\times 20$\,mm\textsuperscript{3}.
The modelled geometry follows that of the Quadra Siemens device~\cite{prenosilPerformanceCharacteristics2022}.

The registered positions, times and energies are smeared using the phenomenological parameterization modelling the experimental resolutions of the Quadra Siemens scanner. The energy resolution dependence is parameterized analogously to \cref{Eq_energy_res}, with the true energy:
$\sigma_i = \frac{2.25}{2.35} \sqrt{E_i}$ (with $E_i$ in keV),
which corresponds to an energy resolution of
10\% \ac{FWHM} at the expected annihilation energy of 511\,keV.
The simulated photon registration time is smeared, event by event, by replacing the event registration time $t_{r}$ by a value obtained from a normal distribution with an energy-dependent standard deviation $\sigma_{t}(E_i)$, corresponding to the scanner \ac{CRT} of 248\,ps at 511\,keV.
In addition, the depth of interaction is modelled by calculating the two intersection points between the photon and the borders of the detector, then uniformly sampling a random position between these two points.
The crystal center closest to the random position is then stored in the resulting list-mode data file.

We simulated a set of 21 point-like source positions within the scanner, located in $(x,0,0)$ for $x\in\{0,2,\dots,20\}$\,cm and $(0,0,z)$ for $z\in\{0,2,\dots,20\}$\,cm.
We did not simulate source positions for negative $x$ or $z$ coordinates, nor positions where $y\neq 0$, due to the symmetries of the simulated scanner.
For each source position, we generated 500000 registered constellations (triple coincidences). Depth of interaction and experimental smearing of position and energy are included as detailed above.
Only true events are modelled; no inter-detector scattering, nor random coincidences are included.

The \ac{TRIO} reconstruction of the simulated data was carried out in MATLAB~7.14.0 (R2012a), with the \ac{NM} optimization performed using the MATLAB \texttt{fminsearch} function from the Optimization Toolbox with the default parameter settings.
The reference energy-based reconstruction algorithm was also implemented in MATLAB according to the description given~\cite{raczynskiVertex2026}.
The reference time-based reconstruction algorithm was implemented in Python~3.10  with the help of NumPy~2.2.

The quality of the reconstructed vertices
is assessed as follows.
Let $\hat{\mathbf{x}}^{\,n}=(\hat{x}^n,\hat{y}^n,\hat{z}^n)\transpose$ for $n\in\{1,2,\dots,N\}$ be the series of reconstructed vertices for $N$ triple photon coincidences originating from the emission source placed at the position $\mathbf{x}=(x,y,z)\transpose$.
For each source position, the bias and the mean position error are calculated.

The bias is defined as the mean signed distance between the reconstructed vertices and the emission point $\mathbf{x}$ along a given axis. For instance, the bias along the $x$ axis is defined as
\begin{equation}
	b_x(\mathbf{x})=\frac{1}{N}\sum_{n=1}^N\left(x-\hat{{x}}^n\right).
	\label{Eq_bias}
\end{equation}
Biases along $y$ and $z$ are defined following the same logic. 

The mean position error, denoted hereafter with  $\varepsilon$, is defined as the mean distance between reconstructed vertices and the emission point. It is defined as:
\begin{equation}
	\varepsilon(\mathbf{x})=\frac{1}{N}\sum_{n=1}^N
	\left\|\mathbf{x}-\hat{\mathbf{x}}^{n}\right\|
\end{equation}

\section{Results}
\label{sec:results}

To  perform  the  comparative  studies  of  the  TRIO algorithm, the analytical trilateration method used in the work of Gajos et al.~\cite{gajosTrilaterationbasedReconstructionOrthopositronium2017}, denoted hereafter with \TimeReco{},
and the energy-based algorithm proposed by Kacperski et al.~\cite{kacperskiThreegammaAnnihilationImaging2004}, denoted hereafter with \EnergyReco{},  were implemented.
Neither the proposed model nor the reference methods require additional parameters or other kinds of optimisation.
The event position is reconstructed directly based on the input data as described in \cref{sec:ortho-reco}.

\begin{figure}
	\centering
	\subfloat[\label{FigErrorRad}]{\includegraphics[width=0.48\textwidth]{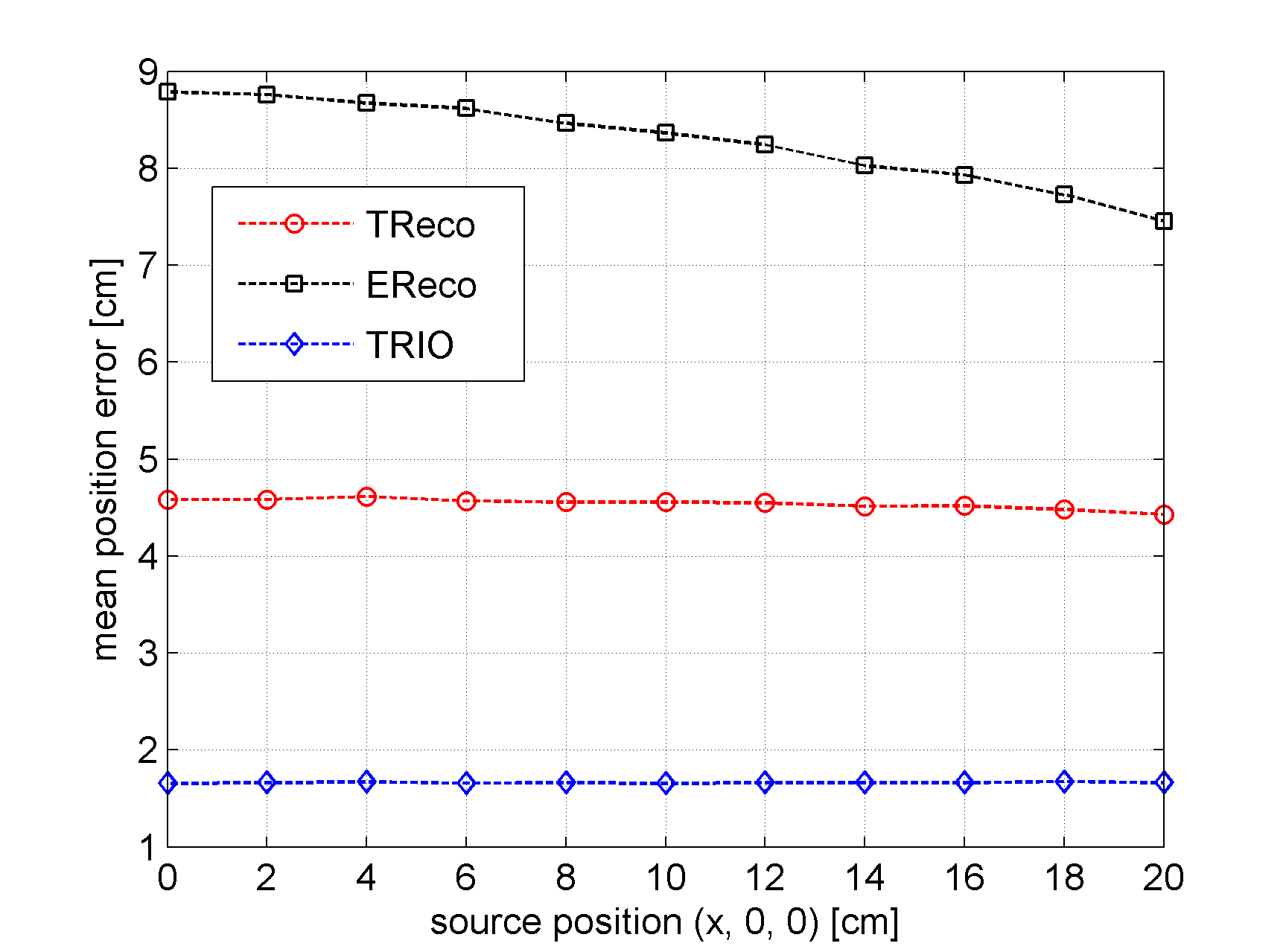}}
	\hfill
	\subfloat[\label{FigErrorAx}]{\includegraphics[width=0.48\textwidth]{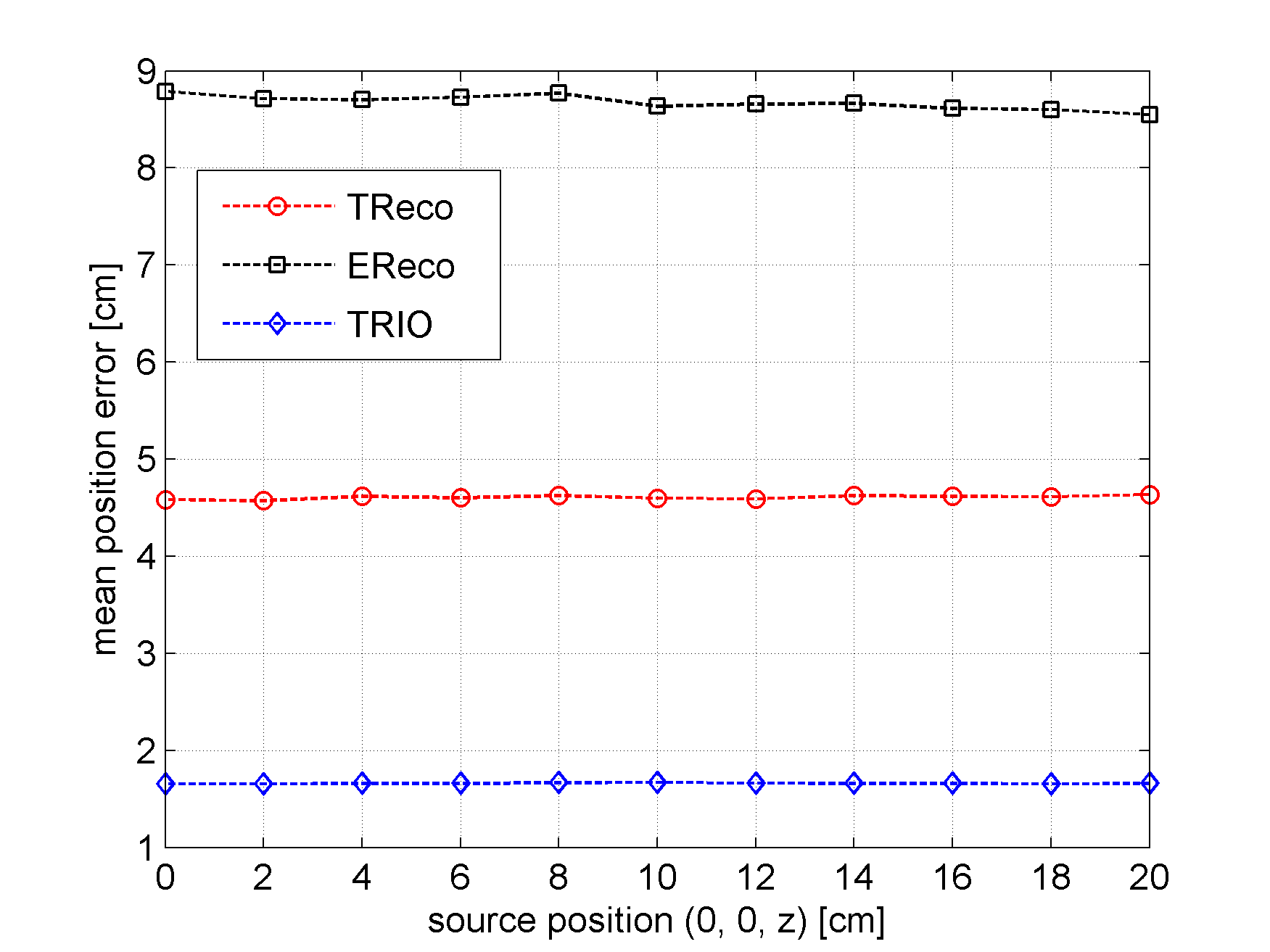}}
	\caption{Mean position errors ($\varepsilon$) for three reconstruction algorithms for point sources along radial (a) and axial (b) direction.}
	\label{Fig:mean_error}
\end{figure}
\Cref{Fig:mean_error} compares the mean position error for three reconstruction algorithms for point sources along radial  (a) and axial  (b) directions.
In the case of the TRIO algorithm,
the $\varepsilon$ reached the value of 1.65~cm on average and  is about 2.8 times smaller in comparison to the \TimeReco{} reconstruction of about 4.59~cm.
The worst results are observed for the
\EnergyReco{} algorithm due to the energy
resolution of the Quadra scanner of about 10$\%$. In this case, the $\varepsilon \approx$~8.7~cm and is about 5.3 times larger when compared to the proposed approach.
It should be underlined that this result is in line with previous observations reported by~\cite{kacperskiThreegammaAnnihilationImaging2004}.

\begin{figure}
	\centering
	\subfloat[\label{FigBiasRad}]{\includegraphics[width=0.48\textwidth]{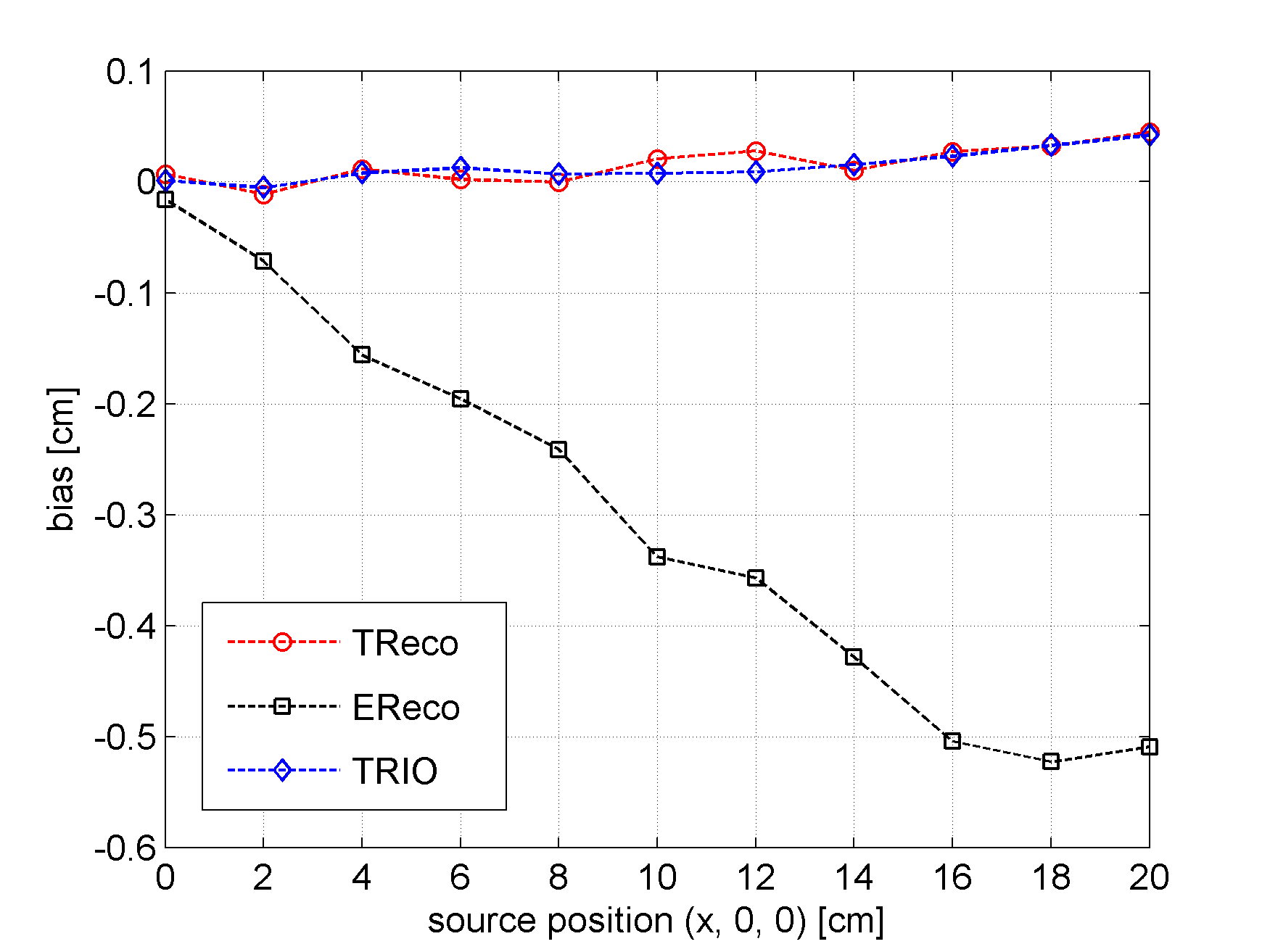}}
	\hfill
	\subfloat[\label{FigBiasAx}]{\includegraphics[width=0.48\textwidth]{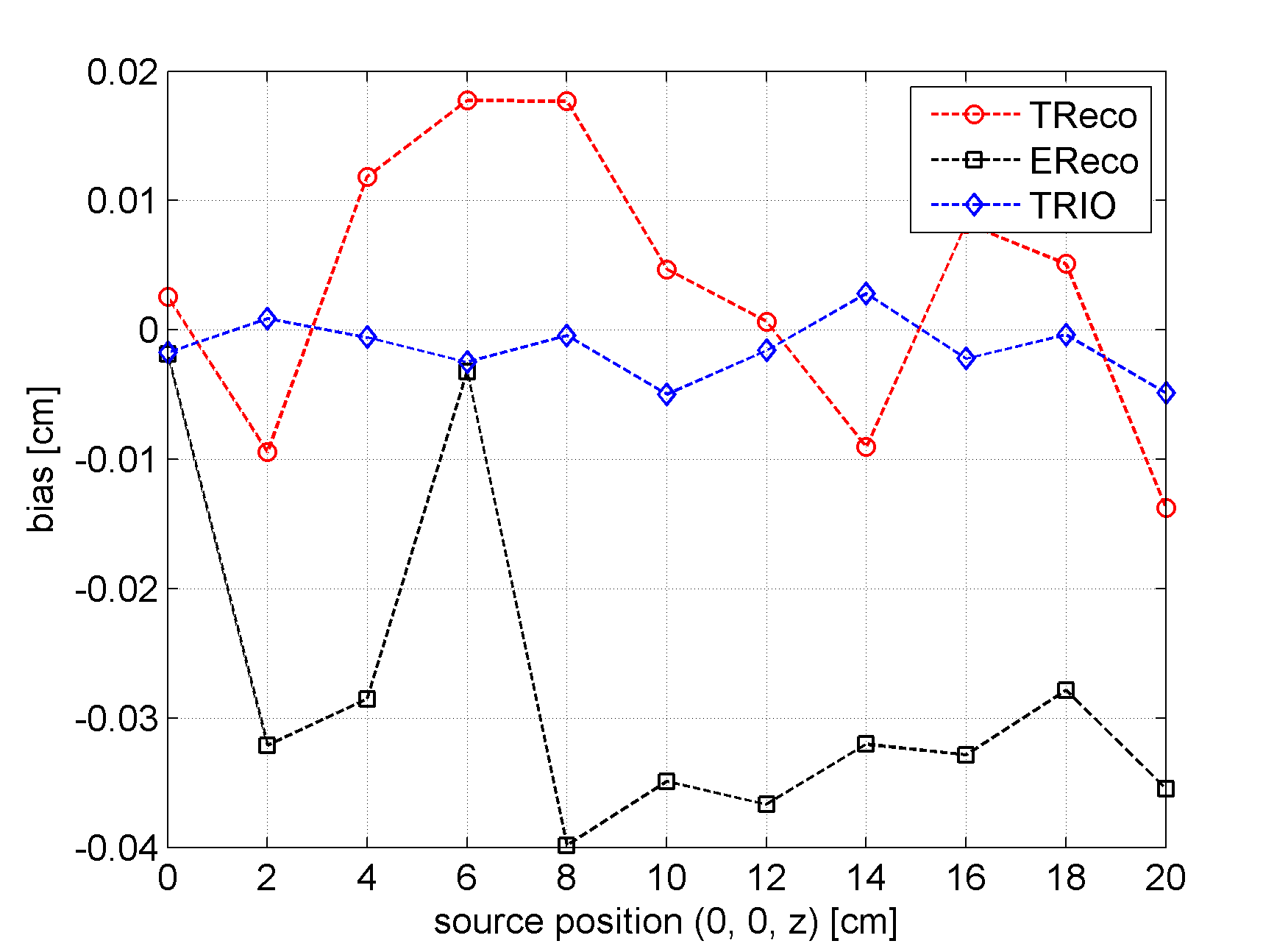}}
	\caption{Biases for three reconstruction algorithms for point sources along radial (a) and axial (b) direction.}
	\label{Fig:bias}
\end{figure}
In \cref{Fig:bias}, the bias of the position reconstruction for three algorithms, for point sources along radial (a) and axial (b) direction, is shown.
In case of the TRIO algorithm and \TimeReco{} reconstruction, the bias
is at a negligible level; in all cases, the absolute value of the bias is lower than 0.05~cm.
On the other hand, the highest absolute values of the bias for \EnergyReco{} are at the level of 0.52 and 0.04~cm for radial and axial directions, respectively.
However, as mentioned, in this case, the performance of the event reconstruction is highly limited by the  energy resolution of the Quadra detector,
which affects both the mean position error and bias.

\begin{figure}
	\centering
	\begin{subfigure}{0.49\textwidth}
		\caption{Prior distribution  $P(\mathbf{x}).$}
		\includegraphics[width=\textwidth]{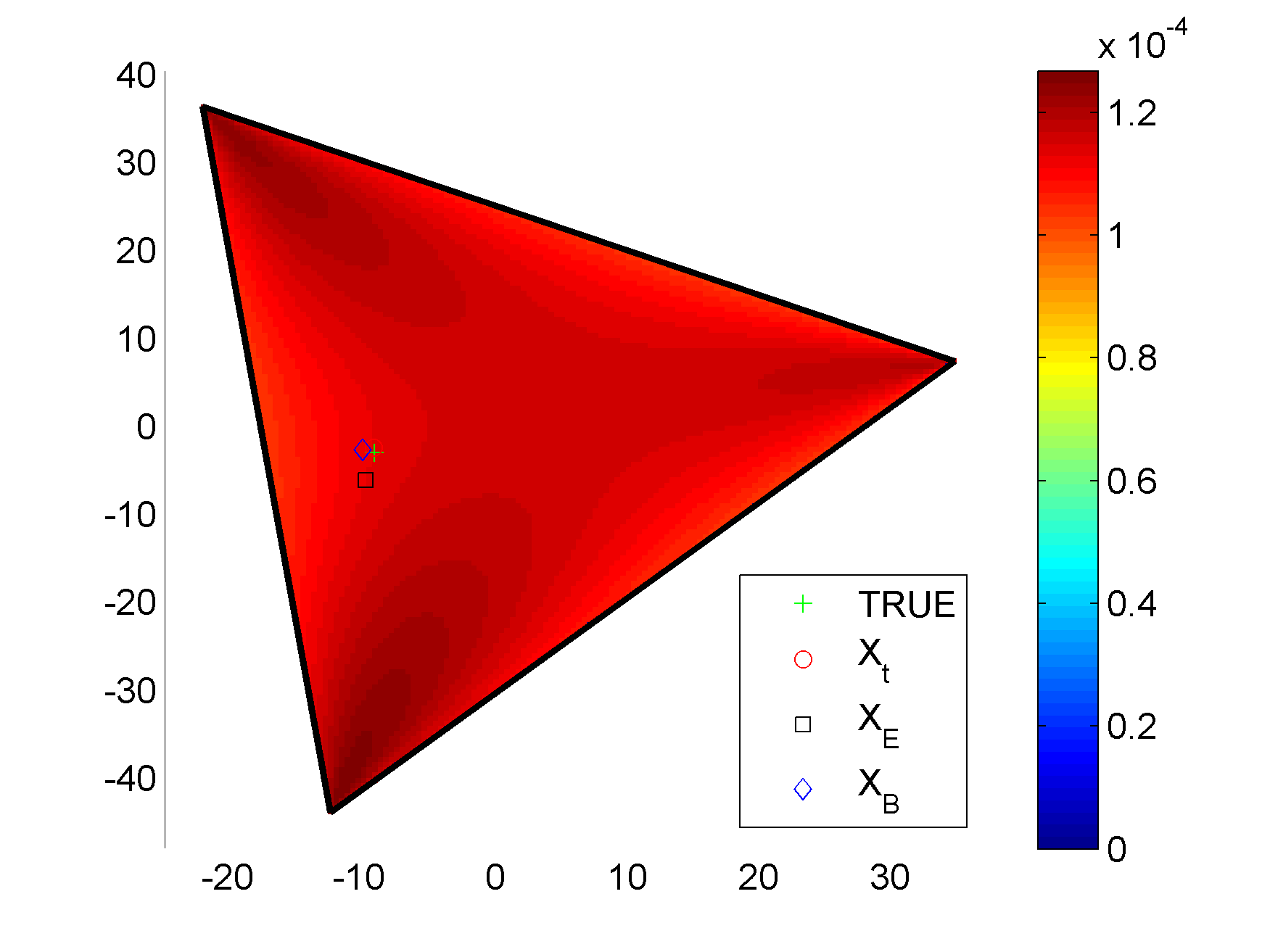}
	\end{subfigure}
	\begin{subfigure}{0.49\textwidth}
		\caption{Probability distribution $P(\mathbf{x_t} | \mathbf{x})$.}
		\includegraphics[width=\textwidth]{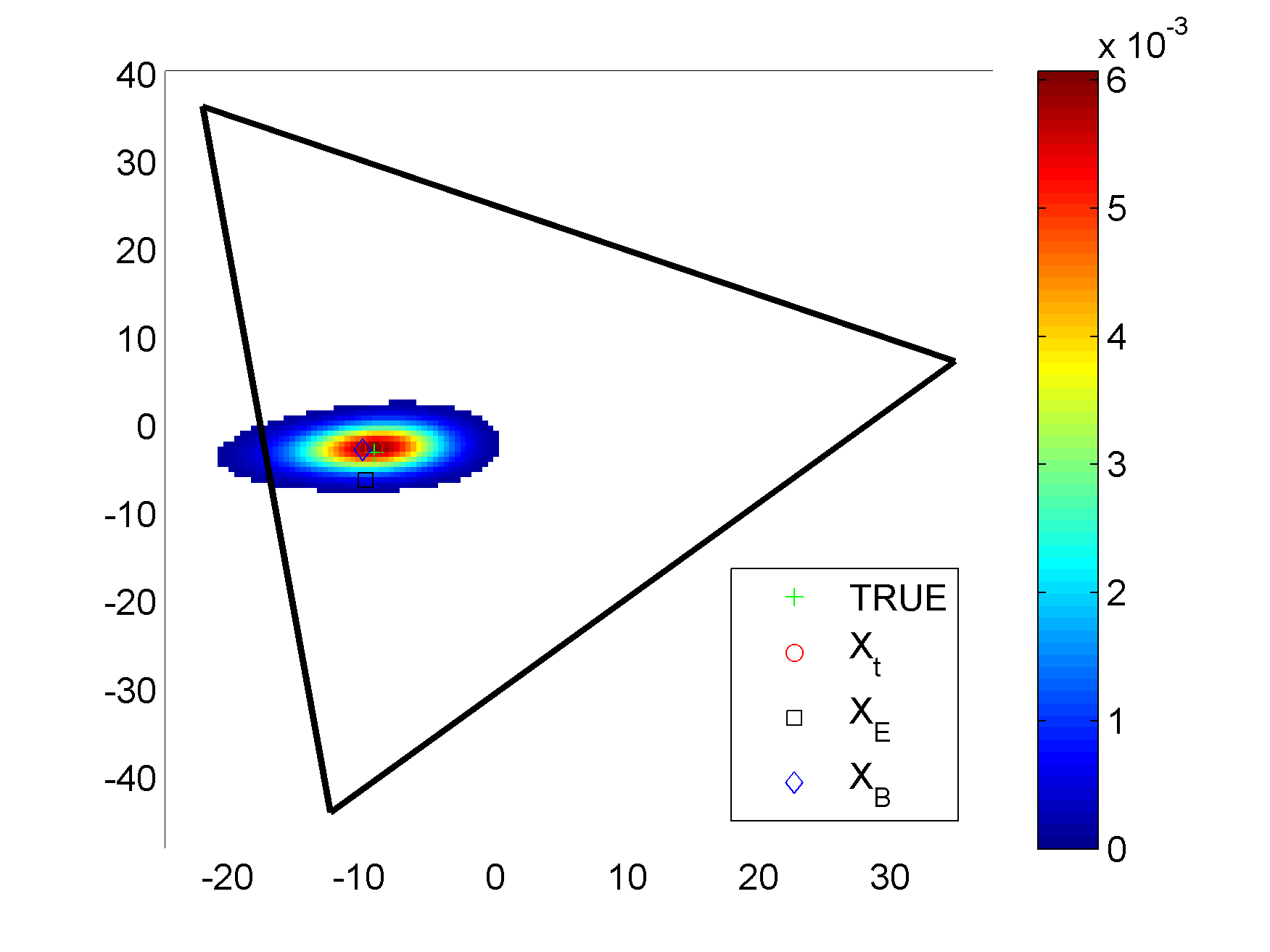}
	\end{subfigure}
	\begin{subfigure}{0.49\textwidth}
		\caption{Probability distribution $P(\mathbf{x_E} | \mathbf{x})$.}
		\includegraphics[width=\textwidth]{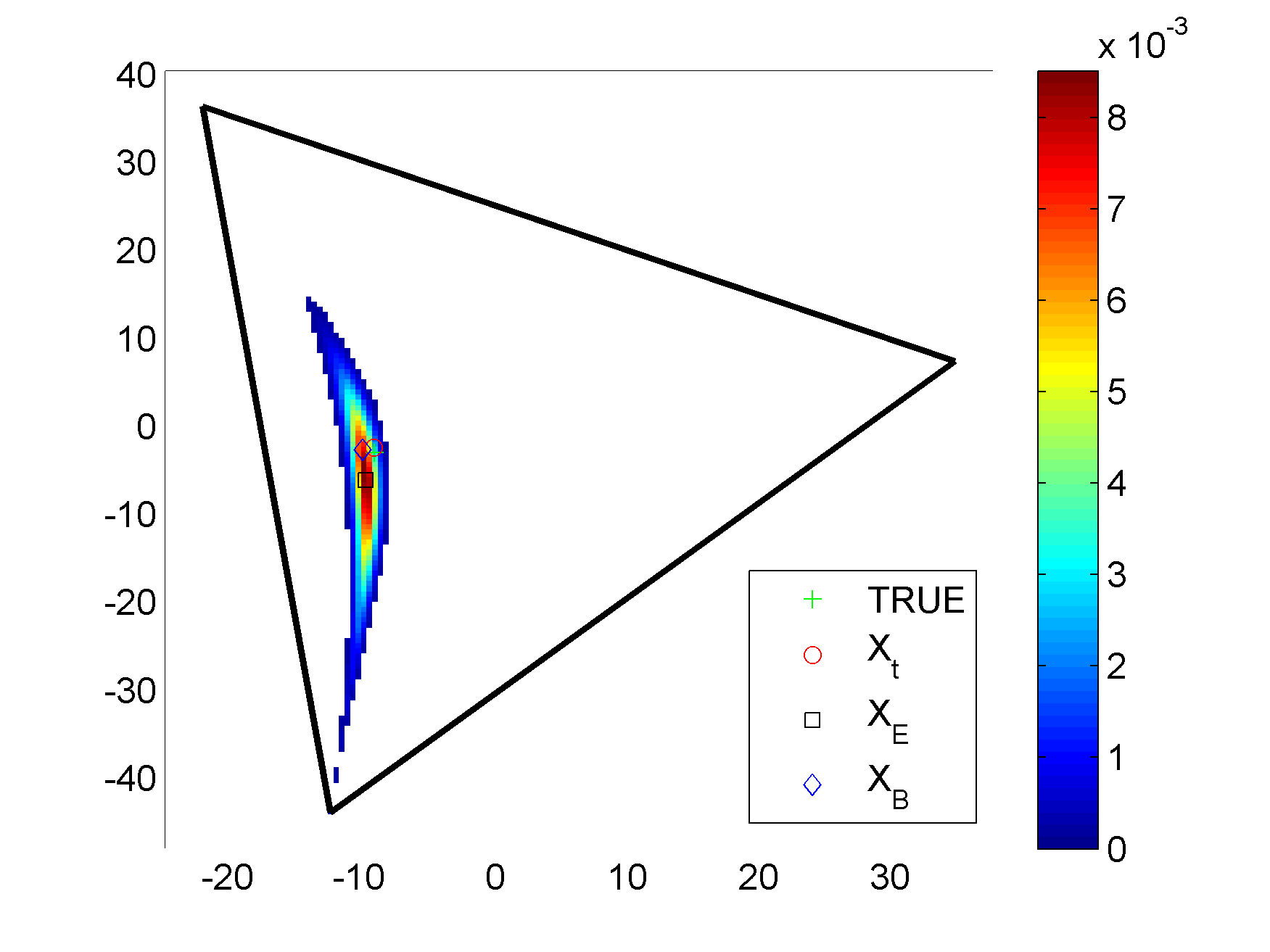}
	\end{subfigure}
	\begin{subfigure}{0.49\textwidth}
		\caption{Probability distribution $P(\mathbf{x} | \mathbf{x_t} \cap \mathbf{x_E})$.}
		\includegraphics[width=\textwidth]{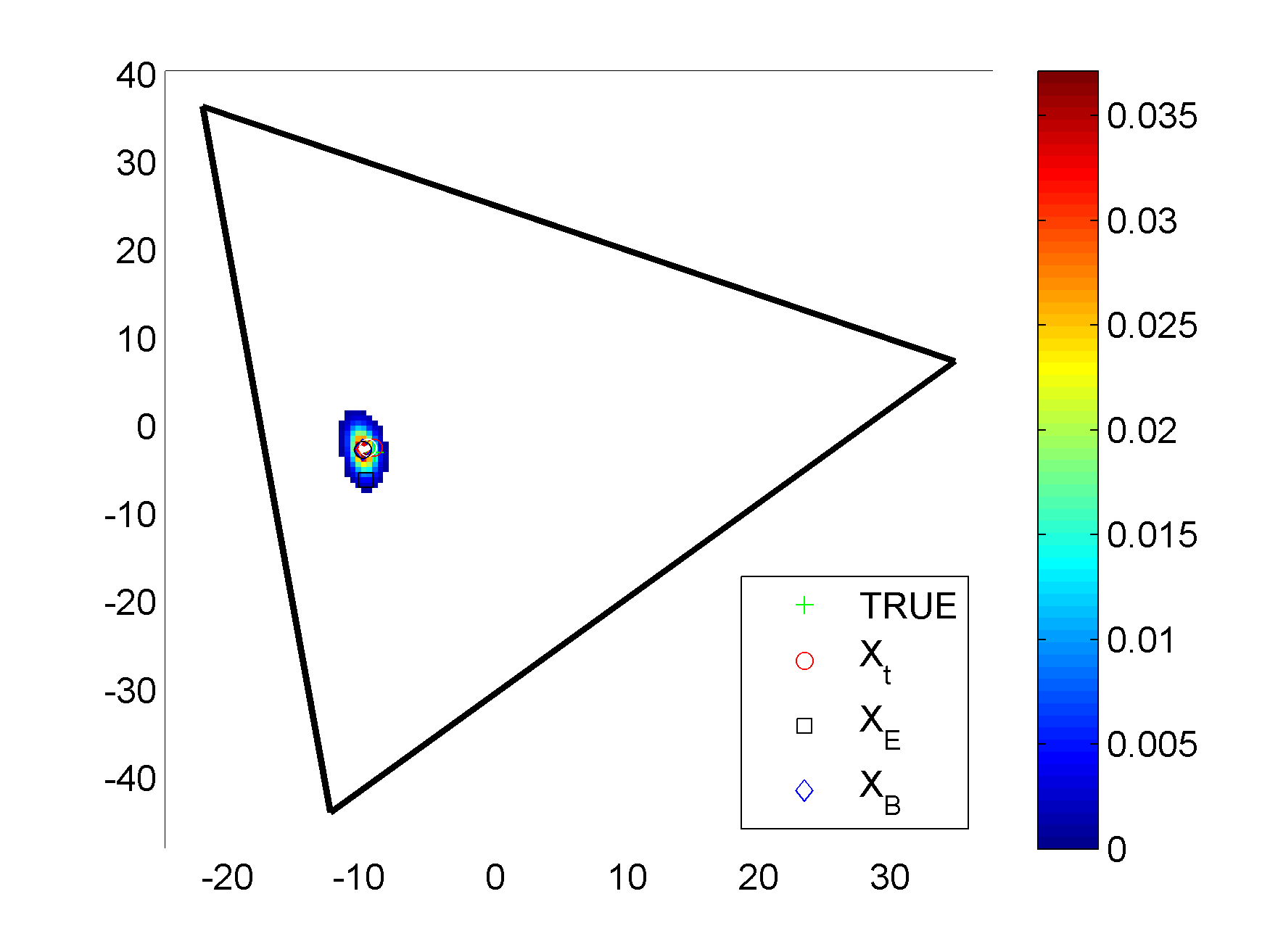}
	\end{subfigure}
	\caption{An example of the position reconstruction of a single \ac{o-Ps} event with additional information about probability distributions on a decay plane.}
	\label{Fig:reco_example}
\end{figure}
An example of the position reconstruction for a single \ac{o-Ps} event using different models is presented in \cref{Fig:reco_example}.
On each panel, the probability distribution calculated on the 2-D decay plane ($x,y$) corresponding to a given method is shown.
Additionally, the distribution of prior probability  $P(\mathbf{x})$, that is used in the TRIO algorithm, is presented in the top left panel in   \cref{Fig:reco_example}.
Three positions of the interactions of the photons with the PET scanner are located at the vertices of the black triangle.
Note that the black triangle defines the regions of feasible solutions with respect to the constraints on the energies of three registered photons.
Therefore, the probability of position reconstruction outside the black triangle is by definition equal to 0 in the case of all methods except \TimeReco{}, which uses only the position and time information.
The reconstructed positions ($\mathbf{x_t}, \mathbf{x_E}$ and $\mathbf{x_B}$) shown in \cref{Fig:reco_example} correspond to the maximal values of the probability distributions
$P(\mathbf{x_t} | \mathbf{x}), P(\mathbf{x_E} | \mathbf{x}),$ and $P(\mathbf{x} | \mathbf{x_t} \cap \mathbf{x_E}),$ respectively, as discussed in \cref{sec:ortho-reco}
(see \cref{Eq_Bayes_xA,Eq_Bayes_xB,Eq_Bayes_xAxB}).
As shown in \cref{Fig:mean_error}, among three reconstruction methods, the energy-based approach is the least accurate.
Despite the fact that the reconstruction $\mathbf{x_E}$ itself is burdened with a large error, the \ac{PDF}  $P(\mathbf{x_E} | \mathbf{x})$
introduces significant information into the statistical model proposed in \cref{Eq_Bayes_xAxB}.
The distribution $P(\mathbf{x_E} | \mathbf{x})$ is stretched and narrow and takes high values also near the true position.
As seen in an example in \cref{Fig:reco_example}, uncertainties of positions $\mathbf{x_E}$ and $\mathbf{x_t}$ are the highest along the vertical and horizontal directions, respectively.
As a consequence, the probability $P(\mathbf{x_E} | \mathbf{x})$ provides a significant improvement to the calculation of the  conditional probability $P(\mathbf{x} | \mathbf{x_t} \cap \mathbf{x_E});$
the final mean position error is about 2.8 times smaller in comparison to the time-based reconstruction ($\mathbf{x_t}$).
Therefore, the application of the proposed Bayesian approach,  where the estimate $\mathbf{x_E}$ is not used directly, but instead all  the distribution $P(\mathbf{x_E} | \mathbf{x})$ is considered,
is beneficial.

In the last part of this work,  the computational speed of
reconstruction algorithms was compared. The efficiency was estimated
only for the models implemented in  MATLAB, that is, \EnergyReco{} and TRIO methods. However, the computational time of the \TimeReco{} algorithm is expected to be similar to the energy-based method, as both reconstructions have analytic solutions.
For comparison purposes, the same total number of 1000 events was considered.
On a single CPU (Intel Core i5-5200U  $@$ 2.20 GHz),
the execution time
of a single event using the TRIO method was  61~ms  on
average.
The evaluation time depends on the overall number of iterations of the \ac{NM} optimization.
We found empirically that, on average, about 40 iterations are required to converge.
On the same CPU, the computing time for the energy-based reconstruction method was  1~ms on average. Hence, reconstruction time in the energy-based method was approximately 61 times shorter than required
by the TRIO model.
It is important to note that the TRIO reconstruction is performed independently for each event, making it naturally parallelizable and well-suited to multicore or GPU-based implementations. Moreover, the current MATLAB implementation is not optimized for performance, and further speed-up is expected after the refactorisation.

\section{Discussion}
\label{sec:discussion}

The TRIO algorithm was compared against two reference methods: time-based trilateration and energy-based reconstruction, using MC simulations of a scanner geometry modelled after the state-of-the-art Siemens Biograph Vision Quadra system. The results demonstrate that TRIO achieves a mean position error of 1.65~cm, representing an improvement of a factor of about 2.8 over the time-based method alone (4.59~cm) and a factor of about 5.3 over the energy-based method (8.7~cm). This improvement reflects the complementary nature of the two likelihood terms, whose product in the Bayesian posterior constrains the solution in both directions simultaneously, yielding a substantially sharper localization than either term alone.
A distinguishing feature of TRIO with respect to all previously proposed three-photon reconstruction methods is the explicit incorporation of a physics-informed prior
derived from the QED description of \ac{o-Ps}  decay.
All previous approaches implicitly assume that the three photons are emitted isotropically.
However, the QED matrix element for three-photon \ac{o-Ps} decay imposes correlations between photon energies and emission angles (see \cref{Eq_Prob_E_priori}), which makes certain decay configurations more probable than others. 
The constraints introduced by the prior are independent of the scanner energy or time smearing parameters. This suggests that its relative contribution would rise as scanner performance deteriorates.

A key advantage of the Bayesian framework is that it naturally adapts to this relative quality of the available measurements. In the Biograph Quadra system, timing resolution is of very good quality (248 ps CRT), while energy resolution, in view of the current analysis, is modest (10\% at 511 keV). Even under these conditions, where the \EnergyReco{} reconstruction alone is highly inaccurate, the energy likelihood still contributes meaningful information to the posterior, as its distribution remains narrow and, in many cases, elongated in the direction orthogonal to the timing uncertainty. This suggests that the Bayesian combination is beneficial even when one of the input modalities is of limited quality, and that TRIO is well suited to the realistic trade-offs encountered in scanner design. For instance, scintillator materials with excellent timing properties (such as LSO or LYSO) often have moderate energy resolution, while detectors optimized for energy resolution (such as HPGe) tend to have poorer timing. The proposed framework accommodates both scenarios gracefully, without requiring re-optimization of the algorithm.

The factorization of the joint likelihood in \cref{Eq_P_x_xt_xE_1} relies on the assumption that $\mathbf{x_E}$ and $\mathbf{x_t}$ are conditionally independent given the true decay position $\mathbf{x}$. Since both estimators share the same measured hit positions $\mathbf{x}_i$, this assumption requires justification. Taking into account the crystal size of $3.2\times 3.2\times 20$\,mm\textsuperscript{3}, the hit position uncertainties are sufficiently small in comparison to the errors introduced by the time and energy measurements to be omitted.
This was verified in our simulations, which include the depth-of-interaction (DOI) modelling and experimental positions smearing. No systematic discrepancies due to a violation of this assumption
are observed. Nevertheless, for scanners with significantly worse spatial resolution or detectors with large DOI uncertainties, both the conditional independence assumption and the coplanarity condition underlying 3D-to-2D transformation should be revisited, and a more general joint likelihood model may be required.

Several aspects of the simulation setup represent simplifications relative to a realistic clinical acquisition. First, only true coincidences are modelled. Neither random nor scatter coincidences are included. These are the dominant sources of background in clinical PET and are expected to degrade the reconstruction accuracy, in particular by introducing spurious triple coincidences. Dedicated event selection strategies, analogous to those used in standard PET for scatter and random corrections, would be required.
The extension to non-point-like activity distributions requires accumulating many reconstructed events and binning them into a volumetric image, which is straightforward in principle but raises additional questions about statistical regularization and image noise. Also, the different strategy might be appropriate if the objective is to achieve the image formed of the three-to-two photon yield ratio. Those questions are out of the scope of the current article, and will be addressed in future works.

Beyond the specific scanner configuration studied here, TRIO applies to any TOF-PET system capable of detecting sub-511 keV photons and registering three-photon coincidences. The overall statistics of the triple coincidences are inherently lower than for standard two-photon annihilations, mainly due to the low event rate in materials, where \ac{o-Ps} decay is dominated by the pick-off and spin conversion processes. As a consequence, only about 0.5\% of all positronium decays correspond to the three-photon events~\cite{kacperskiPerformanceThreephotonPET2005}.  However, as already argued by Kacperski et al. ~\cite{kacperskiPerformanceThreephotonPET2005}, even a moderate statistics sample of three-photon events would allow for the meaningful 3-to-2 imaging.  This limitation can be further overcome by increasing the detection efficiency and event spatial resolution.  Large axial field-of-view systems such as the Quadra or the EXPLORER scanner are therefore particularly well suited, as their high sensitivity partly compensates for this statistical penalty. Furthermore, since TRIO does not require a prompt photon, it is compatible with standard radionuclides such as $^{18}$F-FDG, removing the radionuclide constraint that currently limits \ac{PLI} to specialized tracers such as
$^{44}$Sc. Finally, it is worth noting that the proposed framework is not limited to medical imaging: the event-by-event vertex reconstruction enabled by TRIO could also benefit novel industrial PET techniques as well as fundamental physics studies of \ac{o-Ps}  decay, including tests of discrete symmetries~\cite{allenMeasurementPositroniumDecays2025, moskalTestingCPTSymmetry2021b}. In the latter case, the typical setup uses a well-localized source, and TRIO vertex reconstruction can be used as a selection criterion to discriminate the potential source of the background formed from spurious triple coincidences.

\section{Conclusion}
\label{sec:conclusion}
We have presented TRIO, a novel event-by-event reconstruction algorithm for ortho-positronium three-photon annihilation in PET. TRIO is formulated as a Bayesian maximum a posteriori inference problem that unifies, for the first time, time-based trilateration, energy-based reconstruction, and a physics-informed prior derived from the QED description of \ac{o-Ps} decay within a single probabilistic framework.

Monte Carlo simulations modelled after the Siemens Biograph Quadra scanner demonstrate that TRIO achieves a mean position error of 1.65~cm, improving by a factor of about 2.8 over state-of-the-art time-based trilateration and by a factor of about 5.3 over energy-based reconstruction alone.
The improvement comes from the complementarity of the timing and energy uncertainty in the decay plane, which the Bayesian posterior exploits simultaneously. Crucially, this gain is obtained under the resolution characteristics of current clinical scanners, demonstrating the practical viability of the proposed framework.

Compared to positronium lifetime imaging, TRIO does not require a prompt photon and is therefore compatible with standard radionuclides such as $^{18}$F, broadening its potential clinical applications.

The current study is limited to point sources and true coincidences. Future work will address the extension to volumetric sources, the incorporation of attenuation corrections, and spourious coincidence rejection, and algorithmic acceleration to reduce the per-event reconstruction time. The proposed framework is also applicable beyond medical imaging, offering a tool for novel industrial imaging as well as for fundamental studies of \ac{o-Ps}  physics.

\section*{Acknowledgment}
The authors acknowledge the support by the Foundation for Polish Science through the FIRST TEAM FENG.02.02-IP.05-0152/23 programme co-financed by the European Union under the European Funds for Smart Economy 2021-2027 (FENG). The work is co-financed by the Polish National Agency for Academic Exchange in the frame of the projects BPN/BFR/2025/1/00036/U/00001 and BPN/BAT/2025/1/00009 and co-supported by the Austrian Ministry for Women, Science and Research (project WTZ PL 11/2026).
This work was partially supported by the “PHC POLONIUM” program (project number: 55194PB), funded by the French Ministry for Europe and Foreign Affairs, the French Ministry for Higher Education, Research and Space, and the Polish NAWA.
This work was completed with resources provided by the Świerk Computing Centre at the National Centre for Nuclear Research.

\printbibliography

\end{document}